\documentclass[ALICE,manyauthors]{cernphprep}
\usepackage[comma,square,numbers,sort&compress]{natbib}
\usepackage{hyperref}
\usepackage{lineno}
\usepackage{xspace}
\usepackage{multirow}
\usepackage{graphicx}
\usepackage[loose]{units}
\usepackage{upgreek}
\usepackage{csquotes}
\usepackage{enumerate}
\usepackage{enumitem}
\usepackage{amsmath,amssymb,amsbsy,mathrsfs} 
\usepackage[T1]{fontenc}
\usepackage{orcidlink}

\begin{document}
%

\newcommand{\pp}             {pp\xspace}
\newcommand{\ppbar}        {\mbox{$\mathrm {p\overline{p}}$}\xspace}
\newcommand{\XeXe}         {\mbox{Xe$-$Xe}\xspace}
\newcommand{\PbPb}         {\mbox{Pb$-$Pb}\xspace}
\newcommand{\pA}             {\mbox{pA}\xspace}
\newcommand{\pPb}           {\mbox{p$-$Pb}\xspace}
\newcommand{\AuAu}         {\mbox{Au$-$Au}\xspace}
\newcommand{\dAu}           {\mbox{d$-$Au}\xspace}

\newcommand{\s}                 {\ensuremath{\sqrt{s}}\xspace}
\newcommand{\snn}             {\ensuremath{\sqrt{s_{\mathrm{NN}}}}\xspace}
\newcommand{\pt}                {\ensuremath{p_{\rm T}}\xspace}
\newcommand{\xt}                 {\ensuremath{x_{\rm T}}\xspace}
\newcommand{\meanpt}       {$\langle p_{\mathrm{T}}\rangle$\xspace}
\newcommand{\ycms}           {\ensuremath{y_{\rm CMS}}\xspace}
\newcommand{\ylab}             {\ensuremath{y_{\rm lab}}\xspace}
\newcommand{\etarange}[1] {\mbox{$\left | \eta \right |~<~#1$}}
\newcommand{\yrange}[1]    {\mbox{$\left | y \right |~<~#1$}}
\newcommand{\dndy}           {\ensuremath{\mathrm{d}N_\mathrm{ch}/\mathrm{d}y}\xspace}
\newcommand{\dndeta}        {\ensuremath{\mathrm{d}N_\mathrm{ch}/\mathrm{d}\eta}\xspace}
\newcommand{\avdndeta}    {\ensuremath{\langle\dndeta\rangle}\xspace}
\newcommand{\dNdy}          {\ensuremath{\mathrm{d}N_\mathrm{ch}/\mathrm{d}y}\xspace}
\newcommand{\Npart}          {\ensuremath{N_\mathrm{part}}\xspace}
\newcommand{\Ncoll}           {\ensuremath{\langle N_\mathrm{coll}\rangle}\xspace}
\newcommand{\dEdx}           {\ensuremath{\textrm{d}E/\textrm{d}x}\xspace}
\newcommand{\RpPb}          {\ensuremath{R_{\rm pPb}}\xspace}

\newcommand{\snineH}     {$\sqrt{s}~=~0.9$~Te\kern-.1emV\xspace}
\newcommand{\sseven}     {$\sqrt{s}~=~7$~Te\kern-.1emV\xspace}
\newcommand{\stwoH}      {$\sqrt{s}~=~0.2$~Te\kern-.1emV\xspace}
\newcommand{\stwo}         {$\sqrt{s}~=~2.76$~Te\kern-.1emV\xspace}
\newcommand{\sfive}         {$\sqrt{s}~=~5.02$~Te\kern-.1emV\xspace}
\newcommand{\seight}         {$\sqrt{s}~=~8$~Te\kern-.1emV\xspace}
\newcommand{\sthirteen}   {$\sqrt{s}~=~13$~Te\kern-.1emV\xspace}
\newcommand{\snntwo}     {$\sqrt{s_{\mathrm{NN}}}~=~2.76$~Te\kern-.1emV\xspace}
\newcommand{\snnfive}     {$\sqrt{s_{\mathrm{NN}}}~=~5.02$~Te\kern-.1emV\xspace}
\newcommand{\snneight}     {$\sqrt{s_{\mathrm{NN}}}~=~8.16$~Te\kern-.1emV\xspace}
\newcommand{\LT}             {L{\'e}vy-Tsallis\xspace}
\newcommand{\GeVc}        {Ge\kern-.1emV/$c$\xspace}
\newcommand{\MeVc}        {Me\kern-.1emV/$c$\xspace}
\newcommand{\TeV}           {Te\kern-.1emV\xspace}
\newcommand{\GeV}          {Ge\kern-.1emV\xspace}
\newcommand{\MeV}          {Me\kern-.1emV\xspace}
\newcommand{\GeVmass}  {Ge\kern-.2emV/$c^2$\xspace}
\newcommand{\MeVmass} {Me\kern-.2emV/$c^2$\xspace}
\newcommand{\lumi}           {\ensuremath{\mathcal{L}}\xspace}

\newcommand{\ITS}           {\rm{ITS}\xspace}
\newcommand{\TOF}          {\rm{TOF}\xspace}
\newcommand{\ZDC}          {\rm{ZDC}\xspace}
\newcommand{\ZDCs}        {\rm{ZDCs}\xspace}
\newcommand{\ZNA}          {\rm{ZNA}\xspace}
\newcommand{\ZNC}          {\rm{ZNC}\xspace}
\newcommand{\SPD}          {\rm{SPD}\xspace}
\newcommand{\SDD}          {\rm{SDD}\xspace}
\newcommand{\SSD}          {\rm{SSD}\xspace}
\newcommand{\TPC}          {\rm{TPC}\xspace}
\newcommand{\TRD}          {\rm{TRD}\xspace}
\newcommand{\VZERO}     {\rm{V0}\xspace}
\newcommand{\VZEROA}   {\rm{V0A}\xspace}
\newcommand{\VZEROC}   {\rm{V0C}\xspace}
\newcommand{\Vdecay} 	   {\ensuremath{V^{0}}\xspace}
\newcommand{\EMCal}       {\rm{EMCal}\xspace}
\newcommand{\DCal}          {\rm{DCal}\xspace}

\newcommand{\ee}            {\ensuremath{e^{+}e^{-}}} 
\newcommand{\piz}           {\ensuremath{\pi^{0}}\xspace}
\newcommand{\zz}            {Z\ensuremath{^{0}}\xspace}
\newcommand{\ww}          {W\ensuremath{^{\pm}}\xspace}
\newcommand{\pip}           {\ensuremath{\pi^{+}}\xspace}
\newcommand{\pim}          {\ensuremath{\pi^{-}}\xspace}
\newcommand{\kap}          {\ensuremath{\rm{K}^{+}}\xspace}
\newcommand{\kam}         {\ensuremath{\rm{K}^{-}}\xspace}
\newcommand{\pbar}         {\ensuremath{\rm\overline{p}}\xspace}
\newcommand{\kzero}       {\ensuremath{{\rm K}^{0}_{\rm{S}}}\xspace}
\newcommand{\lmb}          {\ensuremath{\Lambda}\xspace}
\newcommand{\almb}        {\ensuremath{\overline{\Lambda}}\xspace}
\newcommand{\Om}          {\ensuremath{\Omega^-}\xspace}
\newcommand{\Mo}           {\ensuremath{\overline{\Omega}^+}\xspace}
\newcommand{\X}              {\ensuremath{\Xi^-}\xspace}
\newcommand{\Ix}             {\ensuremath{\overline{\Xi}^+}\xspace}
\newcommand{\Xis}           {\ensuremath{\Xi^{\pm}}\xspace}
\newcommand{\Oms}        {\ensuremath{\Omega^{\pm}}\xspace}
\newcommand{\degree}    {\ensuremath{^{\rm o}}\xspace}
\newcommand{\gpro}        {\ensuremath{\gamma^{\rm ~prompt}}\xspace}
\newcommand{\gdec}       {\ensuremath{\gamma^{\rm ~decay}}\xspace}
\newcommand{\gfra}         {\ensuremath{\gamma^{\rm ~frag.}}\xspace}

\newcommand{\ipqcd}  {\ensuremath{I_{\rm pQCD}\xspace}}
\newcommand{\iaa}  {\ensuremath{I_{\rm AA}}\xspace}
\newcommand{\icp}  {\ensuremath{I_{\rm CP}}\xspace}
\newcommand{\raa}  {\ensuremath{R_{\rm AA}}\xspace}
\newcommand{\rcp}  {\ensuremath{R_{\rm CSP}}\xspace}
\newcommand{\rcs}  {\ensuremath{R_{\rm CSC}}\xspace}
\newcommand{\ptg}  {\ensuremath{p_{\rm T}^{\gamma}}\xspace}
\newcommand{\xtg}   {\ensuremath{x_{\rm T}^{\gamma}}\xspace}
\newcommand{\zt}   {\ensuremath{z_{\rm T}}\xspace}
\newcommand{\Dzt}   {D(\ensuremath{z_{\rm T}})\xspace}
\newcommand{\etag}  {\ensuremath{\eta^{\gamma}}\xspace}
\newcommand{\xe}       {\ensuremath{x_{\rm E}}\xspace}
\newcommand{\sigmalong}{\ensuremath{\sigma_{\rm long}^{2}}\xspace}
\newcommand{\sigmalongPb}{\ensuremath{\sigma_{\rm long,~5\times5}^{2}}\xspace}
\newcommand{\tr}{L1-\ensuremath{\gamma}\xspace}
\newcommand{\Deltaphi}{$\Delta\varphi$}

\newcommand{\evt} {\ensuremath{\N_{\rm evt}}\xspace}
\newcommand{\lint}{\ensuremath{\mathcal{L}_{\rm int}}\xspace}

\newcommand{\alphaf}{\ensuremath{\alpha_{\rm MC}}\xspace}
\newcommand{\ptIsoChUE} {\ensuremath{p_{\rm T}^{\rm iso,~UE}}\xspace}
\newcommand{\ptIsoUE}      {\ensuremath{p_{\rm T}^{\rm iso,~ch,~UE}}\xspace}
\newcommand{\ptIsoCh}      {\ensuremath{p_{\rm T}^{\rm iso,~ch}}\xspace}
\newcommand{\ptIso}          {\ensuremath{p_{\rm T}^{\rm iso}}\xspace}
\newcommand{\ptT}   {\ensuremath{p_{\rm T}^{\rm trig}}\xspace}
\newcommand{\ptA}   {\ensuremath{p_{\rm T}^{\rm assoc}}\xspace}
\newcommand{\ptH}   {\ensuremath{p_{\rm T}^{\rm h}}\xspace}
\newcommand{\etaH}  {\ensuremath{\eta^{\rm h}}\xspace}
\newcommand{\pttra}  {\ensuremath{p_{\rm T}^{\rm track}}\xspace}
\newcommand{\etatra}  {\ensuremath{\eta^{\rm track}}\xspace}

\newcommand {\mom}   {\mbox{\rm  GeV$\kern-0.15em /\kern-0.12em c$}}
\newcommand {\gmom} {\mbox{\rm  GeV$\kern-0.15em /\kern-0.12em c$}}
\newcommand {\mass} {\mbox{\rm  GeV$\kern-0.15em /\kern-0.12em c^2$}}
\newcommand{\slfrac}[2]{\left.#1\right/#2}

\begin{titlepage}
\PHyear{2026}       
\PHnumber{127}      
\PHdate{20 April}  

\title{Measurement of isolated-prompt photon$-$hadron correlations in Pb$-$Pb collisions at $\mathbf{\sqrt{\textit{s}_{\textup{NN}}} = 5.02}$~Te\kern-.1emV}
\ShortTitle{Isolated-prompt $\gamma-$hadron correlation in Pb$-$Pb col. at \snnfive }   

\Collaboration{ALICE Collaboration\thanks{See Appendix~\ref{app:collab} for the list of collaboration members}}
\ShortAuthor{ALICE Collaboration} 

\begin{abstract}

The ALICE Collaboration has measured the azimuthal correlation between trigger isolated-prompt photons and associated charged hadrons in Pb$-$Pb collisions at the CERN LHC, at a centre-of-mass energy per nucleon pair of $\sqrt{s_{\mathrm{NN}}}~=~5.02$~Te\kern-.1emV. 
The trigger isolated-prompt photons are measured in the transverse-momentum range $18<p_{\rm T}^{\gamma}<40$~Ge\kern-.1emV/$c$ and pseudorapidity range $|\eta^{\gamma}| <0.67$. 
The isolation selection is based on a charged particle isolation momentum threshold $p_{\rm T}^{\rm iso,~ch} = 1.5$~Ge\kern-.1emV/$c$ within a cone of radius $R=0.2$.
The associated charged particles are measured in the transverse-momentum ranges $p_{\rm T}^{\rm h} > 1.8$~Ge\kern-.1emV/$c$ and pseudorapidity $|\eta^{\rm h}| <0.9$. 
The yield D($z_{\rm T}$) of associated hadrons per trigger, with $z_{\rm T} = p_{\rm T}^{\rm h}/p_{\rm T}^{\rm \gamma}$, is measured in three Pb$-$Pb collision centrality classes: central (0$-$30\%), semicentral (30$-$50\%), and peripheral (50$-$90\%). 
The D($z_{\rm T}$) measurement in pp collisions with the same $p_{\rm T}^{\gamma}$ interval was too limited statistically, but at a lower $p_{\rm T}^{\gamma}$ interval, showing agreement with theoretical predictions. Therefore, an approximation to the standard $I_{\rm AA}$ is computed from the D($z_{\rm T}$) conditional yields, using NLO pQCD predictions as pp reference, the $I_{\rm pQCD}$.
A strong suppression of this ratio is observed in central collisions compared to peripheral collisions.
The result extends to a lower $p_{\rm T}^{\gamma}$ relative to those reported in previously published Pb$-$Pb collisions measurements at $\sqrt{s_{\mathrm{NN}}}~=~5.02$~Te\kern-.1emV.
The $I_{\rm pQCD}$ measurement is compared to $I_{\rm AA}$ calculations with an NLO pQCD model that includes energy loss, and to the CoLBT-hydro model. The $I_{\rm pQCD}$ results from central collisions are also compared with $I_{\rm AA}$ measurements of jets correlated with isolated-prompt photons and of hadrons correlated with Z$^{0}$ bosons, both reported by the CMS Collaboration at the LHC, as well as with direct photon$-$hadron correlation $I_{\rm AA}$ measurements reported by the PHENIX and STAR Collaborations at RHIC.

\end{abstract}
\end{titlepage}

\setcounter{page}{2} 

%
%
\section{Introduction}

Heavy-ion collisions (AA) at ultrarelativistic energies produce a quark--gluon plasma (QGP)~\cite{ALICE:2022wpn, Jacak:2012dx, LHC1review, Braun-Munzinger:2015hba, TheBigPicture, PhenixQGP, StarQGP, PhobosQGP, BrahmsQGP}, a state of deconfined quarks and gluons. 
The high-energy quarks and gluons produced by partonic hard scatterings, which occur at the early stages of the collision, lose energy via collisional and radiative processes in the presence of a QGP~\cite{Djordjevic:2006tw,Ilic:2022bgm}.
As a consequence, the high transverse-momentum (\pt) jet and hadron production, as well as the jet-fragmentation pattern, are modified with respect to their characteristics in proton$-$proton (\pp) collisions: this effect is known as ``jet quenching''~\cite{Bjorken:1982tu, Gyulassy:1999ig, dEnterria:2009xfs}. 
The RHIC and LHC experiments have indeed reported a strong suppression of the production of jets and hadrons for $\pt \gtrsim 5$~\GeVc in central \PbPb and Au$-$Au collisions, which has been attributed to jet quenching~\cite{Adams_2003, PhysRevLett.101.232301, PhysRevC.87.034911, ALICE:2018vuu, Khachatryan_2017, ALICE:2019hno, ALICE:2015mjv, ALICE:2019qyj, ATLAS:2019108, PhysRevC.96.015202}. 

The jet-radiation pattern can be explored via measurements of the fragmentation functions, which are the distributions of the number of constituents within a jet according to their relative contribution to the overall jet transverse momentum.
The ratio of the fragmentation functions in AA over \pp collisions, \iaa, is used to study the jet modifications induced by the QGP, for example, in jet or di-hadron azimuthal correlation measurements~\cite{ATLAS:2018bvp, CMS:2014jjt, STAR:2003pjh, ALICE:2011gpa, ALICE:2016gso}.

Electroweak bosons -- direct-prompt photons ($\gamma$), \zz, and \ww -- 
are produced in the first instants of the collision, do not interact strongly with the QGP, therefore, their production, aside from cold nuclear effects, is unmodified~\cite{Arleo_2007, Aad2016PbPb, Arleo:2011gc, ZHANG_2009}, as reported at RHIC~\cite{PhysRevLett.94.232301, Afanasiev:2012dg} and the LHC~\cite{Aad:2019sfe,  Chatrchyan:2012nt, ALICE:2022cxs,Aad:2012ew, Chatrchyan:2014csa,Aad:2019lan, PhysRevLett.127.102002, Afanasiev:2012dg, Chatrchyan:2012vq, Sirunyan:2020, ALICE:2024yvg, ALICE-PUBLIC-2024-003}.
At leading order (LO) in perturbative quantum chromodynamics (pQCD), they are produced in association with a parton emitted back-to-back in azimuthal angle and with similar \pt, which, in contrast to the boson, interacts with the QGP.
Therefore, electroweak bosons can be used as a reference for the energy scale of the hard scattering process -- as such, they are called ``triggers''.
The distribution of hadrons from the hadronisation of the parton recoiling from the trigger can be expressed with the following proxy for the parton fragmentation function:  $\Dzt = \frac{1}{N^{\rm trig}} \frac{{\rm d}N^{\rm h}}{{\rm d}\zt}$, with $\zt = \ptH/ \pt^{\rm trig}$, 
where $N^{\rm trig}$ is the number of trigger particles, \ptH and $\pt^{\rm trig}$ are the charged particle and trigger transverse momenta, respectively, and $N^{\rm h}$ is the number of charged particles with momentum fraction \zt\ associated to the trigger. 
Azimuthal-correlation measurements between electroweak bosons and hadrons provide an effective way to probe the nuclear modification of the parton fragmentation function, avoiding biases in jet-energy reconstruction present in reduced-jet areas, which can be modified by jet quenching. 
Direct $\gamma-$hadron correlations were measured at RHIC by the STAR~\cite{STAR:2016jdz} and PHENIX~\cite{PHENIX:2012aba} Collaborations in Au$-$Au collisions at $\snn=200$~GeV, and direct $\gamma-$jet~\cite{CMS:2018mqn,ATLAS:2019dsv} and Z$^{0}-$hadron~\cite{ATLAS:2020wmg,CMS:2021otx} correlations were measured at the LHC by the CMS and ATLAS Collaborations in \PbPb\ collisions at \snnfive. 
In those measurements, a strong modification of the jet pattern with respect to \pp collisions has been observed. 
The ALICE Collaboration also measured isolated-prompt $\gamma-$hadron correlations in \pp\ and \pPb\ collisions at \snnfive~\cite{ALICE:2020atx}, where no modification of the fragmentation pattern was observed, a result expected since QGP formation, and therefore, jet quenching is not likely in such collisions.

The kinematic range probed for \PbPb collisions in the measurement presented in this article, using direct-prompt photons as triggers, measured via the isolation technique described below, offers access to lower $Q^{2}$ relative to other LHC experiments, where the largest nuclear effects can be expected, and to a similar \ptg range as RHIC measurements. The relative energy loss and the steepness of the partonic spectra can enhance the observable suppression, and the interplay between parton virtuality and medium scales may become more relevant.

At LO in pQCD, direct-prompt photons are produced via $2 \rightarrow 2$ processes: (i)
quark--gluon Compton scattering $\rm{q g} \rightarrow \rm{q} \gamma$, and (ii) quark--antiquark annihilation $\rm{q \overline{q}} \rightarrow g \gamma$, with a small contribution from $\rm{q \overline{q}} \rightarrow \gamma \gamma$. 
In addition, non-direct prompt photons are produced by higher-order processes, like parton fragmentation or Bremsstrahlung.
Requiring the photons to be ``isolated'' allows suppression of the contributions not only from fragmentation and Bremsstrahlung photons~\cite{Ichou:2010wc} but also from photons originating from hadron decays, as these photon sources are commonly accompanied by other parton fragments.
The isolation criterion requires that the sum of the transverse momenta of the produced particles (\ptIso) in a cone with angular radius $R$ around the photon direction is smaller than a given threshold value. 
The advantage of this selection is that it can be applied in both experimental measurements and theoretical calculations. 

This paper presents the isolated-prompt $\gamma-$hadron correlations measured in \PbPb collisions at $\sqrt{s_{\rm NN}}=5.02$~TeV  by the ALICE Collaboration, using data samples collected in the years 2015 and 2018. The trigger particles are isolated-photon candidates in a pseudorapidity ($\eta$) range $|\etag|<0.67$ and a \pt range of $18<\ptg<40$~\GeVc, that were selected with an isolation cone radius $R=0.2$ and isolation momentum threshold $\ptIsoCh = 1.5$~\GeVc using only charged particles in the cone. 
The associated charged particles are measured with $\ptH > 1.8$~\GeVc\ and $|\etaH| <0.9$.
The isolated-photon identification and event selections are the same as in the isolated-prompt photon \pt-differential spectra measurement presented in Ref.~\cite{ALICE:2024yvg}.

The correlation procedure followed is close to the one presented for the measurement in \pp and \pPb collisions at \snnfive in Ref.~\cite{ALICE:2020atx}. 
In this article, the \Dzt distribution obtained for hadrons emitted opposite to the photon, at azimuthal angle with respect to the photon of $3/5\pi<|\Delta \varphi| = \varphi^{\gamma}-\varphi^{\rm h}<\pi$~rad, is presented. 
To observe the modification of hadron fragmentation due to jet quenching, a baseline free of nuclear effects is needed. In this measurement, theoretical next-to-leading-order (NLO) pQCD calculations without jet quenching are used. Additionally, the data are confronted with theoretical models that include energy loss.
The ratios of the \Dzt distribution to the NLO pQCD calculation without energy loss are compared to the ratios of the \Dzt in AA over pp collisions, both calculated with theory models including energy loss or measured in different RHIC and LHC experiments.

This paper is divided into the following sections: Section~\ref{sec:detector} briefly presents the detector setup and the data sample used for the analysis; Section~\ref{sec:photonsel} describes the photon selection procedure; Section~\ref{sec:corr} presents the azimuthal correlation and \Dzt\ measurements and corrections. The systematic uncertainties are presented in Sec.~\ref{sec:sys_unc}, and the final results and conclusions are presented in Secs.~\ref{sec:results} and~\ref{sec:conclusion}, respectively.
Additional figures giving more details on this analysis are available in Ref.~\cite{ALICE-PUBLIC-2026-001}.

\section{Detector and event selection\label{sec:detector}}

The ALICE experiment and its performance during the LHC Run 2 (2015--2018) are described in Refs.~\cite{Aamodt:2008zz,Abelev:2014ffa}.
Photon reconstruction was performed using the Electromagnetic Calorimeter (EMCal)~\cite{ALICE:2022qhn}, which covers $\Delta \varphi<173^\circ$ and $|\eta|<0.7$, while charged particles used in the photon isolation and azimuthal correlation were reconstructed with the combination of the Inner Tracking System (ITS)~\cite{Aamodt:2010aa} and the Time Projection Chamber (TPC)~\cite{Alme:2010ke}, with a combined coverage of $|\eta|<0.9$ and full azimuthal angle. The forward scintillator arrays (V0)~\cite{ALICE:2013axi} and zero degree calorimeters (ZDC)~\cite{Abelev:2014ffa} were used for online triggering, and for event selection and characterisation. For a brief description of the detector's setup, see Ref.~\cite{ALICE:2024yvg}.

The data were taken with a minimum bias (MB) interaction trigger and EMCal Level-1 photon-dedicated triggers (\tr). 
The MB trigger is defined as a coincidence between the V0A and the V0C (forward and backwards V0 detectors) trigger signals.
In the 2015 \PbPb sample, the MB triggered data were taken so that the centrality distribution was uniform, but for the 2018 data sample, the 0$-$10\% and 30$-$50\% centrality classes were enhanced with dedicated V0 triggers. 
Events above 90\% centrality are excluded, since there are substantial contributions from electromagnetic processes, and their low multiplicity results in an inefficient trigger.
The \tr\ triggers are based on energy depositions in $4\times4$ calorimeter cells larger than 10~Ge\kern-.1emV in \PbPb for the year 2015. For the 2018 \PbPb collisions, the threshold has been set at 10~Ge\kern-.1emV for the 50\% more central collisions, and at 5~Ge\kern-.1emV for the other centrality classes. 
A detailed description of the \tr\ triggers can be found in Ref.~\cite{ALICE:2022qhn}.

An offline event selection based on the V0 timing information is applied to remove beam-induced background events~\cite{ALICE:2013hur}. In addition, further beam-background reduction is obtained in \PbPb collisions using the information from two ZDCs positioned at 112.5~m on either side of the nominal interaction point. In particular, a selection is applied to the correlation between the sum and the difference of times measured in each of the ZDCs~\cite{Abelev:2014ffa}.
Finally, only events with a primary vertex along the beam direction within $\pm 10$~cm from the centre of the apparatus are considered in this analysis, to grant a uniform acceptance in $\eta$.

The measurement presented here is performed in three \PbPb collision centrality classes:  
0$-$30\%, 30$-$50\%, and 50$-$90\%. The integrated luminosity used for each centrality class are $ 7767\pm 75$~nb$^{-1}$, $1325\pm 21$~nb$^{-1}$ and $378\pm 8$~nb$^{-1}$, respectively, the same as in Refs.~\cite{ALICE:2024yvg, ALICE-PUBLIC-2024-003}. 

\section{Isolated-photon reconstruction and selection}
\label{sec:photonsel}

Isolated-prompt photons are reconstructed via the following steps:
 (a) reconstruction of clusters of cells in the calorimeter and of tracks with the ITS and the TPC;
 (b) photon identification via charged-particle vetoing using track--cluster matching and via the cell-energy spread (shower shape); 
 and (c) selection of isolated-photon candidates. 
This section briefly describes the selection procedure, however, all details can be found in Refs.~\cite{ALICE:2024yvg, ALICE-PUBLIC-2024-003}, where the same data samples are used, and the same selections are applied.
 
In the EMCal, made of independent readout channels called ``cells'', particles deposit their energy in several adjacent cells. 
Each such fragmented energy deposit is grouped into a cluster by a clusterisation algorithm. Various versions of these algorithms, together with the detector calibration procedure and corrections, are described in detail in Ref.~\cite{ALICE:2022qhn}. 
While single photons create almost round clusters, other clusters can have a wider, elongated shape; for example, when several particles deposit their energy nearby in the detector, or when electrons hit the EMCal at a pronounced angle.
The most frequent cases are neutral-meson decays into two photons, which are reconstructed as a single cluster when the angular distance between both photons is such that their electromagnetic showers overlap partially. 
Merged and single-photon clusters can be discriminated by the ``shower shape'' variable \sigmalongPb, which is derived from the largest eigenvalue of the cluster cell spatial distribution in the $\eta-\varphi$ plane~\cite{ALICE:2022qhn} restricted to $5\times$5 cells centred in the cluster highest-energy cell~\cite{ALICE:2024yvg, ALICE-PUBLIC-2024-003}.
In this article, ``photon candidates'' refer to clusters with a ``narrow'' circular shape, i.e., a small shower shape value ($0.1<\sigmalongPb<0.3$), while ``\piz candidates'', which are used to estimate the background in the signal region, refer to clusters with a ``wide'' elliptic shape ($0.4<\sigmalongPb<1$). For more details and discussion, see Refs~\cite{ALICE:2024yvg, ALICE-PUBLIC-2024-003}.

An isolation criterion is applied to the photon candidate to suppress the contribution by fragmentation and neutral-meson decay photon production. 
In this measurement, this is based on the so-called ``isolation momentum'', $\ptIsoCh=\sum p_{\rm T}^{\rm h}-\pi \times R^{2} \times \rho_{\rm UE}$, i.e. the sum of the transverse momenta of all charged particles ($\pt^{\rm h}$), 
inside a cone of radius $R < \sqrt{ (\eta^{\rm h} - \eta^{\gamma})^2 + (\varphi^{\rm h}-\varphi^{\gamma})^2}$ around the photon candidate, located at coordinates $\eta^{\gamma}$ and $\varphi^{\gamma}$ in the angular space, 
and where $\eta^{\rm h}$ and $\varphi^{\rm h}$ are the coordinates of the measured charged particles.
The isolation momentum is corrected by the contribution of the underlying-event (UE) momentum density ($\rho_{\rm UE}$), estimated in a rectangular band around and excluding the isolation cone covering $|\Delta \eta^{\rm h}|<0.9$ and $\Delta \varphi^{\rm h}=2(R+0.1)$ rad (details can be found in Refs.~\cite{ALICE:2024yvg,ALICE-PUBLIC-2024-003}).
Typically, a value of $R<0.4$ is used in \pp and \pPb\  collisions~\cite{ALICE:2019rtd, ALICE:2024kgy, ALICE:2020atx, ALICE:2025bnc}, but the smaller value $R=0.2$ was chosen for this measurement to reach purity and efficiency values as large as possible in central \PbPb collisions, given the large underlying-event contribution in the cone~\cite{ALICE:2024yvg, ALICE-PUBLIC-2024-003}. 
The purity is indeed higher in central collisions for $R = 0.2$ than for $R=0.4$ since the underlying-event fluctuations are significantly smaller and thus less background is accepted as isolated.
Accepted tracks in the cone are required to satisfy $|\eta^{\rm h}|< 0.9$ and $\pt^{\rm h}>0.15$~\GeVc. The track definition is given in Refs.~\cite{ALICE:2024kgy, ALICE-PUBLIC-2024-003}, and is the same as that used for the hadron correlation in the next section.
The candidate photon is declared isolated if $p_{\rm T}^{\rm iso,~ch}<$~1.5\,\GeVc.

Although selections were applied, the isolated-photon candidate sample still contains a sizeable contribution from background clusters, mainly from neutral-meson decay photons. The sample purity is estimated by a procedure known as the ``ABCD method''~\cite{ Aad:2010sp,Aad:2011tw,Aad:2013zba,Aad:2016xcr,Aad:2017,ALICE:2019rtd,ALICE:2024kgy,ALICE:2024yvg, ALICE:2025bnc}, which uses double ratios between four different classes of measured clusters (wide and narrow, isolated and non-isolated) and a Monte Carlo-based factor to correct the correlation between the four classes.
The obtained purity ranges from 0.5 in the least central class to 0.65 in the most central class~\cite{ALICE:2024yvg, ALICE-PUBLIC-2026-001}.

\section{Azimuthal correlation}
\label{sec:corr}

The azimuthal correlation distribution is constructed using isolated-photon candidate triggers with a momentum in the range $18<\ptT<40$~\GeVc, and a pseudorapidity $|\eta^{\rm trig}|<0.67$. They are associated with charged particles, i.e. tracks measured by the TPC and ITS, that are selected with a transverse momentum $\ptH >1.8$~\GeVc and $|\etaH|<0.9$. The choice of the lower \ptT and 
\ptH limits are driven by the large collision underlying event, preventing a clear signal observation in the lowest \zt intervals.

The raw conditional yield of hadrons with respect to the trigger
is written as
  $S(\Delta\eta,|\Delta\varphi|)= \frac{1}{N^{\rm{trig}}}\frac{\text{d}^{2} N^{\text{h}}(\Delta\eta,|\Delta\varphi|)}{\text{d}\Delta\eta\text{d}|\Delta\varphi|} $,
where $\Delta\eta = \eta^{\rm trig} - \etaH$, $|\Delta\varphi| = \varphi^{\rm trig} - \varphi^{\rm{h}}$ (the absolute value indicates that the distribution is mirrored for \Deltaphi$ > \pi$~rad and \Deltaphi$ < 0$ rad to reduce the statistical uncertainties),  $N^{\rm{trig}}$ is the number of trigger particles, and $N^{\rm{h}}$ is the number of associated hadrons.

The analysis strategy consists of four main steps to obtain the isolated-prompt photon \Dzt\ distribution using azimuthal correlations to associated tracks triggered by high-\pt\ isolated-narrow (prompt-$\gamma$ signal + \piz\ background) and isolated-wide (\piz) clusters: 
\begin{enumerate}[noitemsep,topsep=-1pt]
  \itemsep0em
\item estimation and subtraction of the underlying-event contribution from $S(\Delta\eta,|\Delta\varphi|)$ in the collision; 
\item estimation and subtraction of the background contribution in the trigger sample, primarily due to \piz\ decays;
\item for each \zt\ interval, integration of the resulting azimuthal distribution in the $|\Delta\varphi|$ range opposite to the trigger-particle direction;
\item correction for the track-reconstruction efficiency.
\end{enumerate}

The particles originating from the collision's underlying event are uncorrelated to the hard process which produces the trigger particles considered in this measurement. 
The underlying-event contribution to the azimuthal correlation conditional yield is therefore estimated by correlating trigger clusters with tracks from other collisions with similar characteristics.
A pool of such collisions, called the ``mixed-event sample'', is filled for different intervals of: 
centrality, in steps of 10\% between 0 and 90\%; reaction plane angle~\cite{Poskanzer:1998yz}, in steps of $\pi$/4 between 0 and $\pi$~rad; and $z$-vertex position, within intervals of 2~cm between $+10$~cm and $-10$~cm.
The resulting distributions, $M(\Delta\eta,|\Delta\varphi|)$, 
which are obtained similarly to $S(\Delta\eta,|\Delta\varphi|)$ but normalised by the total number of mixed events per candidate trigger used, 
are then subtracted from the same-event distributions,  $C(\Delta\eta,|\Delta\varphi|)~=~S(\Delta\eta,|\Delta\varphi|)-M(\Delta\eta,|\Delta\varphi|)$.

\begin{figure}[ht]
\centering
\includegraphics[width=0.94\textwidth]{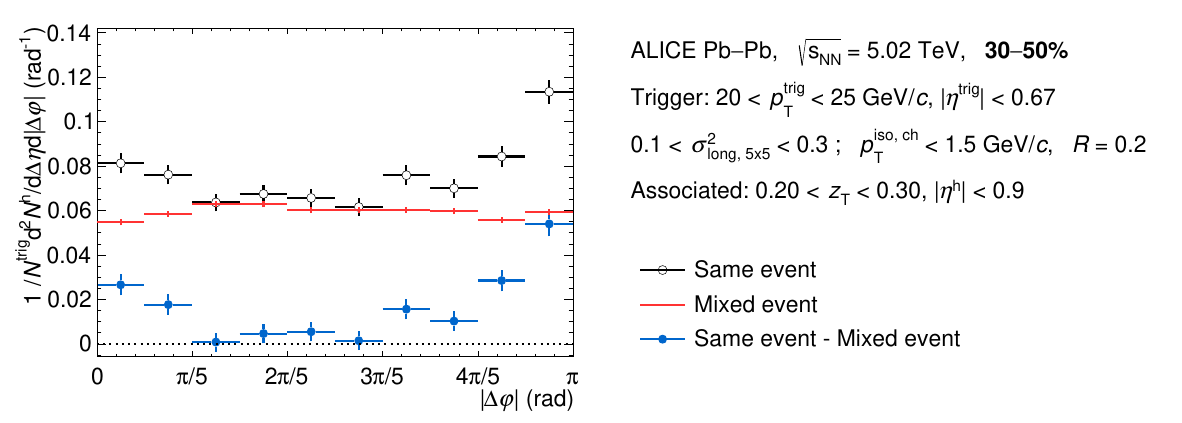} 
\caption{(colour online) 
Azimuthal correlation distributions in the 30$-$50\% centrality class of Pb$-$Pb collisions at $\sqrt{s_{\mathrm{NN}}}~=~5.02$~Te\kern-.1emV and for the interval $0.20<z_{\rm T}<0.30$, for isolated-narrow clusters triggered in the range $20 < p_{\rm T}^{\rm trig}~<25$~Ge\kern-.1emV/$c$, with charged-particle tracks in the same collision (``Same event'', black-open circles) or in a different collision (``Mixed event'', red without marker). The result of subtracting the mixed-event distribution is shown in blue-full circles. 
The vertical bars indicate the statistical uncertainties. 
}
\label{fig:DPhi_samemix}
\end{figure}

Figure~\ref{fig:DPhi_samemix} shows the raw azimuthal yield of hadrons associated with isolated-narrow cluster triggers, for semicentral \PbPb collisions in a given \ptg and \zt interval. Distributions for the same \ptT\ interval but for different centrality and \zt intervals, 
can be found in Ref.~\cite{ALICE-PUBLIC-2026-001}. 
The mixed- and same-event distributions are similar to each other in the angular region perpendicular to the trigger direction ($|\Delta\varphi| \sim \pi/2$), which indeed is expected to be dominated by the underlying event.  
The hard-process contribution is visible as a peak due to the recoiling jet, located on the side opposite to the trigger ($|\Delta\varphi| =\pi$~rad), but the background contamination in the photon candidate sample also generates a peak on the trigger side ($|\Delta\varphi| =0$~rad).

\begin{figure}[ht]
  \begin{center}
  \includegraphics[width=0.94\textwidth]{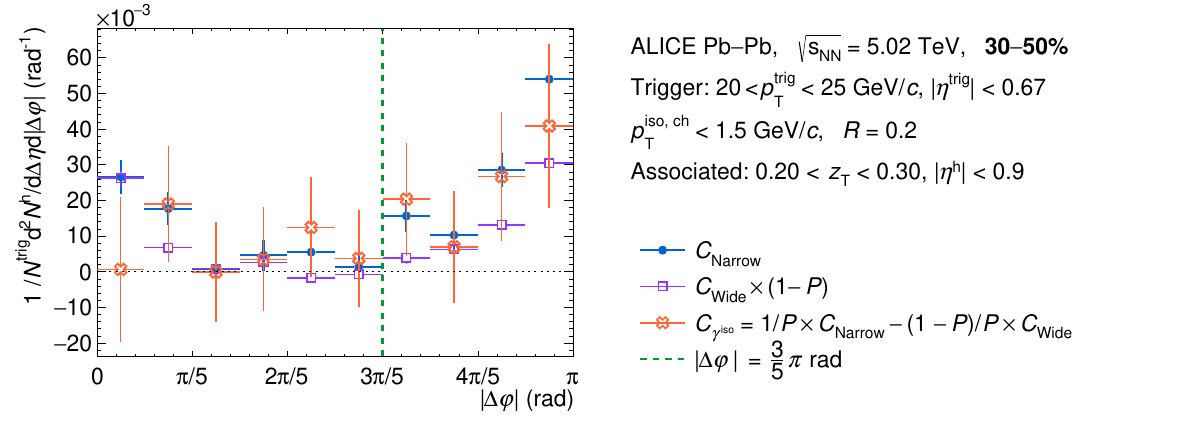} 
  \end{center}
  \caption{\label{fig:DPhi_NarrowWide}
  (colour online) 
  Azimuthal correlation distributions in the 30$-$50\% centrality class of Pb$-$Pb collisions at $\sqrt{s_{\mathrm{NN}}}~=~5.02$~Te\kern-.1emV
  for the intervals $0.20<z_{\rm T}<0.30$ and $20<p_{\rm T}^{\rm trig}<25$~Ge\kern-.1emV/$c$. Each marker corresponds to a different trigger: isolated-narrow clusters (blue-full circles), also shown in Fig.~\ref{fig:DPhi_samemix}; isolated-wide clusters (violet-open squares), scaled by $1-P$, where $P$ is the purity;
  isolated-photon triggers (orange-empty crosses), calculated using the equation in the text.
  The vertical bars indicate the statistical uncertainties. The dashed-green line at $|\Delta \varphi|=3/5\pi$ indicates the lower limit of the integration used in the D($z_{\rm T}$) calculation.} 
\end{figure}

Photons from \piz\ decays represent the largest fraction of the contamination in the $C_{\rm narrow}$ distribution,  
the rest comes mainly from $\eta$ meson decays. 
$\eta$ and \piz\ triggers at the \pt\ range covered by this measurement, significantly larger than their masses, present $C$ distributions that agree in PYTHIA 8 simulations. The $C_{\rm wide}$ is almost entirely triggered by \piz\ decays. 
Therefore, assuming that photon clusters from neutral meson decays with narrow or wide shapes have the same azimuthal correlation distribution, the isolated-prompt $\gamma$--hadron azimuthal correlation is calculated by
$C_{\gamma^{\rm iso}}(\Delta\eta,|\Delta\varphi|) = \frac{1}{P} \times C_{\rm narrow}(\Delta\eta,|\Delta\varphi|) - \frac{1-P}{P} \times C_{\rm wide}(\Delta\eta,|\Delta\varphi|)$, 
where $P$ is the purity presented in the previous section that ranges from 0.5 to 0.65 from the least to the most central class~\cite{ALICE-PUBLIC-2026-001}.
At the \ptT interval of this measurement, background clusters, narrow and wide, are mostly those containing two photons from the meson decay.
The results of this procedure are shown in Fig.~\ref{fig:DPhi_NarrowWide} for semicentral \PbPb~collisions. Figures for the same \ptT\ interval but for different centrality and \zt intervals can be found in Ref.~\cite{ALICE-PUBLIC-2026-001}.
The values of the correlation functions for narrow and wide trigger clusters are very close, resulting in a significant statistical uncertainty in the measurement.

The \Dzt distribution is obtained by integrating the points above 3$\pi$/5, and correcting for geometrical acceptance, detector inefficiencies and resolution, inactive detector areas, and underlying-event subtraction effects.
In order to obtain the correction factor, the detector response is modelled by Monte Carlo simulations reproducing the detector conditions during the data-taking periods.
The corrections are obtained using PYTHIA~8 (version 8.210~\cite{Sjostrand:2014zea} with the Monash 2013 tune~\cite{Skands_2014}) as a particle generator, creating \pp 
collisions in intervals of transverse momentum of the hard scattering with a prompt photon and a jet ($\gamma$--jet) calculated at LO in the final state. 
The transport of the generated particles in the detector material is performed using GEANT3~\cite{Geant3}.   
Each simulated \pp collision is embedded into a real \PbPb minimum-bias triggered event selected within the different centrality classes considered,  
so that the effect of the low-energy particles from the underlying event is properly taken into account.
For the calorimeter clusters, the embedding is performed at the cell level by summing the cell energy of the data and of the simulation. 
For the charged particles, the embedding is done at the track level, adding the tracks coming from the data to the list of available tracks from the simulation.
The \Dzt function from the $\gamma$--jet simulation is obtained in the same way as in data. 
The correction factor is calculated as the ratio of the \Dzt values for generated tracks and reconstructed tracks, 
which is almost constant for all centralities for $\zt>0.2$ and is approximately 1.2. 
For smaller \zt, the correction factor decreases below unity for semicentral and peripheral collisions, but increases for the most central collisions up to a value of almost three (more details in Ref.~\cite{ALICE-PUBLIC-2026-001}).
This article reports in Sec.~\ref{sec:results} the correlation for central collisions in the centrality class 0$-$30\%, which was obtained
by adding the corrected conditional yields of the centrality classes 0$-$10\% and 10$-$30\%, weighted by the number of triggers in each class.

\section{Systematic uncertainties}
\label{sec:sys_unc}

The considered sources of systematic uncertainty are the isolated-photon purity, the shower-shape selection for wide clusters, the tracking efficiency, the centrality intervals chosen to fill the mixed-event pool, and the mixed-event underlying-event estimation. 
The values of the systematic uncertainties for the different sources and the total systematic uncertainty are summarised in Table~\ref{table_systDzt} for two \zt intervals.

The total uncertainty assigned to the purity correction, $\sigma_{P}$, reported in Ref.~\cite{ALICE:2024yvg,ALICE-PUBLIC-2024-003} is used to vary the purity value $\pm \sigma_{P}$, resulting in an uncertainty with a \zt\ dependence, from low to high \zt: 
from 6\% to 20\% in central collisions, from 3\% to 11\% in semicentral collisions, and from 5\% to 16\% in peripheral collisions. 
Semicentral collisions have a lower uncertainty than peripheral collisions due to the higher purity and lower systematic uncertainty.

The uncertainty due to the choice of the background wide-cluster \sigmalongPb\ range is investigated by comparing the results obtained for various \sigmalongPb\ selections: $0.35 <\sigmalongPb< 1.0$,  $0.5 <\sigmalongPb~<~1.0$, and $0.4 <\sigmalongPb< 0.8$. The resulting uncertainty is \zt-independent and changes from 6\% in central collisions to 4\% in peripheral collisions.

To estimate the effect of the tracking efficiency on the measurement, a fraction of randomly chosen tracks is removed. 
This fraction is defined by the tracking efficiency uncertainty due to the matching of the ITS-TPC tracks, 
which varies with track \pt: for central collisions, it is
about 3.6\% at low \pt (below 1 \GeVc), decreasing to about 2\% for larger momenta; for semicentral and peripheral collisions, it has a similar magnitude and track-\pt\ dependence, but is lower by less than 0.5\%.

The effect of the centrality ranges used in the mixed-event pool is checked by selecting intervals two times smaller for the centrality (5\% instead of 10\%): the uncertainty, \zt-independent, is 6\% in central and 4\% in peripheral collisions. For the $z$-vertex and reaction-plane angle, the effect of using finer intervals was found to be negligible.

Finally, an uncertainty is assigned to the hypothesis that the mixed-event method could not fully capture the underlying event. 
Assuming that the $C_{\rm \gamma^{\rm iso}}$ and $C_{\rm wide}$ distributions must be flat and sit at zero in the region $1 < \Delta\varphi < \pi /2$,
the relative systematic uncertainty is calculated as $\frac{\rm B/2}{\rm I-B/2}$,
where $I = \int{N^{\rm{h}}(\Delta\varphi)}\text{d}\Delta\varphi$ is the integral of the $3\pi/5 < \Delta\varphi < \pi$ signal region, and $B = \int{C_{\text{fit}}}\text{d}\Delta\varphi$ is the integral of the constant fit value $C_{\rm{fit}}$ in $1 < \Delta\varphi < \pi /2$ of the  $C_{\rm wide}$ distribution ($C_{\rm \gamma^{\rm iso}}$ has large statistical fluctuations and it is not expected to be different). 
This uncertainty decreases from lowest- to highest-\zt\ interval from 8.8\% to 6\% in central collisions, from 7.3\% to 4.5\% in semicentral collisions, and from 8\% to 6\% in peripheral collisions.

The total systematic uncertainty is obtained by adding in quadrature the contributions from all the sources.
It ranges from 15.7\% to 23.5\% (from low to high \zt) for central collisions, 9.6\% to 14.6\% (minimum at mid \zt and maximum at high \zt)  for semicentral collisions, and 12.7\% to 18.9\% (from low to high \zt) for peripheral collisions.
Among the systematic uncertainty sources, the mixed-event subtraction source dominates in the first \zt\ intervals, where the other sources contribute at the level of 3\% to 6\%; only starting from \zt\ larger than 0.35, the purity uncertainty source dominates. 
The statistical uncertainty dominates the systematic one over the whole \zt\ range in all the \PbPb collision centrality classes. 
It ranges between 20\% and 40\% for most of the points.

\begin{table}[ht!]
\centering
\caption{Summary of uncorrelated relative systematic uncertainties in per cent for two selected high- and low-\zt intervals: $0.10 < \zt < 0.15$ and $0.40 < \zt < 0.60$.} 
\label{table_systDzt}
\begin{tabular}{lcccccc}
\hline
 & \multicolumn{2}{c}{0$-$30\%}
 & \multicolumn{2}{c}{30$-$50\%}
 & \multicolumn{2}{c}{50$-$90\%} \\
\cline{2-7}
 & low $z_{\rm{T}}$ & high $z_{\rm{T}}$ & low $z_{\rm{T}}$ & high $z_{\rm{T}}$ & low $z_{\rm{T}}$ & high $z_{\rm{T}}$\\
\hline
Photon purity&   6.4\%&   20.3\%&   2.7\%&   11.6\%&   5.5\%& 16.6\%   \\
$\sigma_{\rm long,~5\times5}^{2}$&   7.9\%&   7.9\%&   5.8\%&   5.8\%&   5.6\%& 5.6\%   \\
Underlying event&   8.8\%&   5.9\%&   7.2\%&   4.5\%&   8\%& 5.9\%   \\
Mixed-event pool&   6.1\%&   6.1\%&   4.7\%&   4.7\%&   3.9\%& 3.9\%   \\
Tracking efficiency&   5.3\%&   1.7\%&   4.1\%&   1.8\%&   4.4\%& 1.6\%   \\
\hline
Total& 15.7\%& 23.5\%& 11.5\%& 14.6\%& 12.7\%& 18.9\% \\
\hline
\end{tabular}
\end{table}

For the ratio of the measurement in central or semicentral to peripheral collisions, the uncertainty due to the tracking efficiency and the effect of the centrality ranges in the mixed-event pool fully cancels out, and the others partially, except the uncertainty on the mixed-event underlying-event removal, which is fully propagated and dominates. The total systematic uncertainty ranges from 14\% to 16\% in central over peripheral collisions, and from 13\% to 14\% in semicentral over peripheral collisions.

\section{Results\label{sec:results}}

\begin{figure}[!htbp]
      \includegraphics[width=1\textwidth]{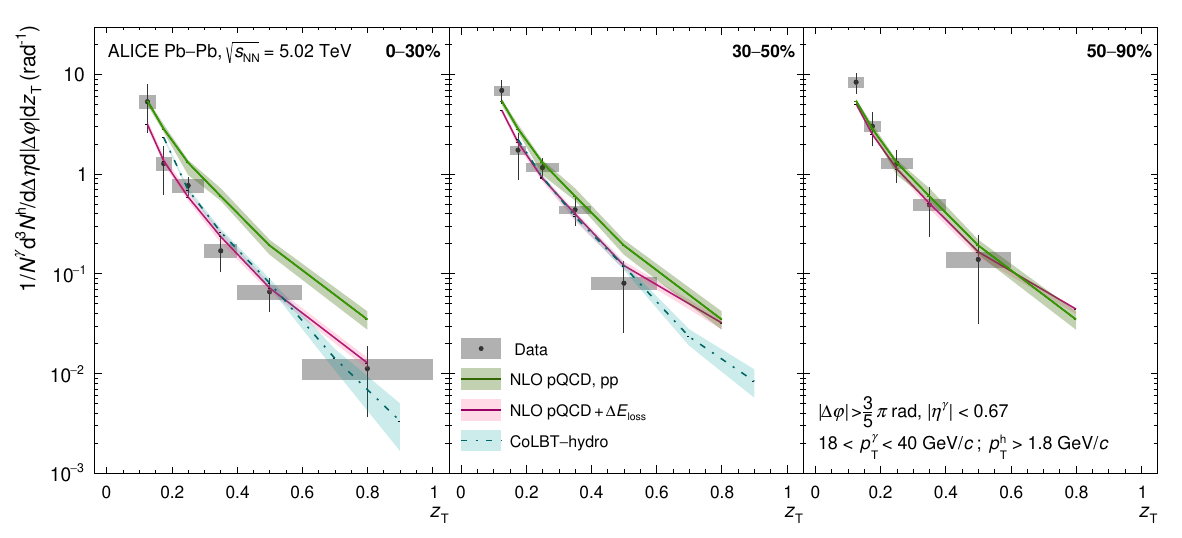}
    \caption{\label{fig:Dzt} (colour online) 
 D($z_{\rm T}$) distributions for isolated-prompt photon$-$hadron correlations measured in Pb$-$Pb collisions at $\sqrt{s_{\mathrm{NN}}}~=~5.02$~Te\kern-.1emV for three centrality classes: 0$-$30\% (left), 30$-$50\% (middle), and 50$-$90\% (right). 
 The boxes and vertical lines represent the systematic and statistical uncertainties, respectively. 
 Different theory predictions are shown: 
 an NLO pQCD calculation including energy loss~\cite{Xie:2020zdb, Zhang:2009rn} (pink line and uncertainty band), and from CoLBT-hydro~\cite{Chen:2017zte} (cyan-dashed line and uncertainty band). Also, an NLO pQCD prediction for pp collisions at $\sqrt{s}~=~5.02$~Te\kern-.1emV is shown (green line and uncertainty band).
 }
\end{figure}

This section presents the main results of the isolated-prompt photon$-$hadron correlation measurement. The isolated-photon trigger is measured in a transverse-momentum range of $18<\ptg<40$~\GeVc at midrapidity ($|\etag| < 0.67$), with a charged-particle isolation-momentum threshold of $\ptIsoCh = 1.5$~\GeVc in a cone of radius $R = 0.2$ around the photon candidate. 
The associated charged particles are measured with $\ptH > 1.8$~\GeVc and $|\etaH| <0.9$. The \Dzt\ distribution is extracted after subtracting the underlying-event and decay-photon contribution, by integrating the azimuthal angle region $|\Delta \varphi|>3\pi/5$. 
Figure~\ref{fig:Dzt} shows the measured \Dzt distributions in three \PbPb collision centrality classes: 0$-$30\%, 30$-$50\%, and 50$-$90\%.  

The results are compared with NLO pQCD calculations including energy loss in the QGP~\cite{Xie:2020zdb, Zhang:2009rn} (``NLO pQCD+$\Delta E_{\rm loss}$'').
They include jet energy loss controlled by the in-medium energy-loss density $\hat{q}$ and medium temperature $T$, and calculated by the Higher-Twist formalism~\cite{Deng:2009ncl} with energy loss given by $\hat{q} / T^3$, extracted from single hadron \pt spectrum, di-hadron, and direct $\gamma-$hadron correlation data at different \snn~energies (0.2, 2.76, and 5.02~TeV) by the IF-Bayesian analysis~\cite{Xie:2022fak, Xie:2022ght, Wang:2003aw}.
The model obtains the parton$-$parton hard scattering cross sections by perturbation calculation and convolves them with the CT18A parton density functions~\cite{Hou:2019efy} and Kniehl-Kramer-Potter fragmentation functions~\cite{Kniehl:2000fe}. 
The NLO pQCD calculations include fragmentation photons, which are reduced by applying the same isolation method.
The calculated uncertainty is based on the extrapolated $\hat{q} / T^3$ energy loss at the 95\% confidence level.

The results in central and semicentral collisions are also compared with the ``Coupled Linear Boltzmann Transport and hydrodynamics'' model~\cite{Chen:2017zte} (``CoLBT-hydro''). 
The CoLBT-hydro model is developed for event-by-event simulations of jet transport and jet-induced medium excitation in high-energy heavy-ion collisions by a (3+1)D hydrodynamics that has a source term from energy-momentum deposition by propagating jet shower partons~\cite{Chen:2017zte}.  
In the CoLBT-hydro model, the interactions are again driven by $\hat{q}$, but its value is extracted mainly from a single jet and isolated $\gamma-$jet correlation measurements in \PbPb collisions at \snnfive. 
The calculation uncertainty is derived from the $\hat{q}$ estimation uncertainty.

Both models use the same centrality classes and kinematical selection criteria for the trigger and associated charged particles as those used in the data.
An agreement is observed between the data and both theoretical predictions. 
The discrimination between the two models is not yet possible due to current uncertainties. Figure~\ref{fig:Dzt} also displays the \Dzt\ distribution obtained from NLO pQCD simulations of \pp collisions, without energy loss. 
The uncertainty in the calculations is estimated by varying the factorisation scale $\mu$ of the fragmentation functions: 0.7\ptg, 1.2\ptg (centre value), and 2\ptg.
The measured \Dzt\ distributions are suppressed compared to the  NLO pQCD calculations for \pp collisions for central and semicentral \PbPb collisions, reflecting the high-\pt\ hadron suppression in the presence of the QGP. 

\begin{figure}[!htbp]
    \begin{center}
    \includegraphics[width=1\textwidth]{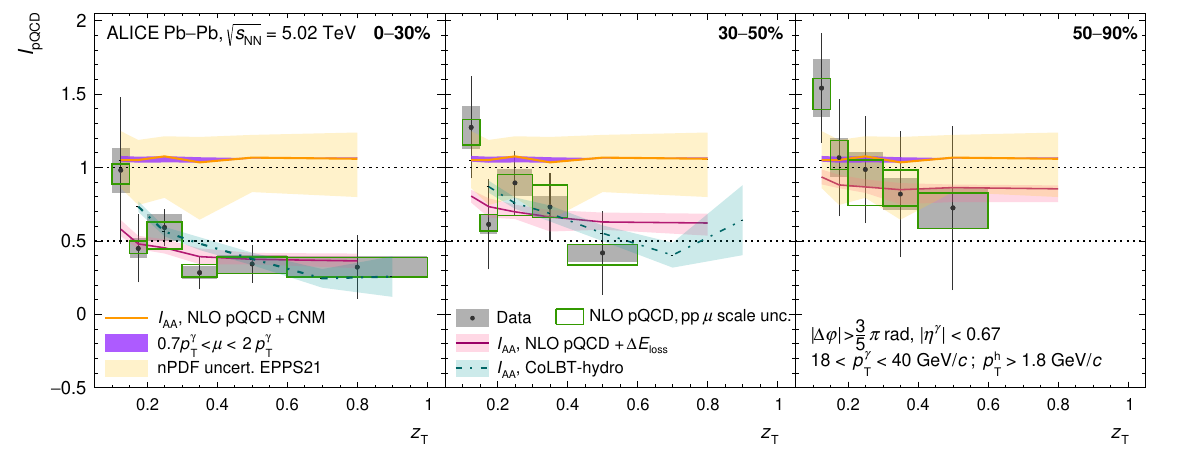}
    \end{center}
    \caption{\label{fig:Ipqcd} (colour online) Ratio of the measured D($z_{\rm T}$) distributions in Pb$-$Pb collisions at $\sqrt{s_{\mathrm{NN}}}~=~5.02$~Te\kern-.1emV to the NLO pQCD predictions for pp collisions at $\sqrt{s}~=~5.02$~Te\kern-.1emV, $I_{\rm pQCD}$, all distributions in the denominator and numerator are shown in Fig.~\ref{fig:Dzt}. 
    Each panel shows a different centrality class: 0$-$30\% (left), 30$-$50\% (middle), and 50$-$90\% (right).
    The grey-filled boxes and vertical-black lines represent the data systematic and statistical uncertainties, respectively. 
    The green-open boxes represent the $\mu$ scale uncertainty for the denominator NLO pQCD calculation for pp collisions. 
    Different theory predictions for the $I_{\rm AA}$, which include energy loss, are shown: an NLO pQCD calculation~\cite{Xie:2020zdb, Zhang:2009rn} (pink line and uncertainty band), and from a CoLBT-hydro calculation~\cite{Chen:2017zte}  (cyan-dashed line and uncertainty band, not for peripheral collisions). Also, an NLO pQCD prediction, which includes CNM effects only (no energy loss, no centrality dependence), is reported (light orange line and uncertainty bands). 
    }
\end{figure}

To better illustrate the modification, the ratio of the \Dzt distributions in AA to \pp collisions, the $\iaa~=~\frac{\Dzt_{\text{AA}}}{\Dzt_{\text{pp}}}$, is calculated in per-trigger yield measurements.
The \Dzt\ distribution for isolated $\gamma-$hadron correlations has been measured by the ALICE Collaboration in \pp\ collisions at $\sqrt{s} = 5.02 $ TeV~\cite{ALICE:2020atx}, but with a different \ptg~range ($12<~\ptg~<40$~\GeVc). 
Due to the large amount of underlying events present in the \PbPb collisions, it is not possible to go as low in \ptg as in the \pp collisions measurement. On the other hand, a reanalysis of the \pp collisions data using a higher \ptg threshold is not accessible due to the statistical limitations on that sample. 
For these reasons, the NLO pQCD calculations (no energy loss, no cold nuclear matter effects) are used as a pp collisions reference, which have been shown to describe the pp collisions data well. 
 In Ref.~\cite{ALICE-PUBLIC-2026-001}, a comparison between the published pp results and this pQCD calculation, performed over a different  \ptg\ interval, is presented. The results are found to be consistent with the calculation, taking into account the expected 5$-$10\% deviation in PYTHIA 8 arising from the difference in the \ptg\ interval.
Therefore, the NLO pQCD calculations are used in this measurement as a proxy of the $\iaa$ denominator, by using the ratio $\ipqcd= \frac{\Dzt_{\text{AA}}}{\Dzt_{\text{pQCD pp}}}$, which is shown in Fig.~\ref{fig:Ipqcd}. 

Excluding the first \zt interval, a clear difference can be seen in central collision data with respect to the NLO pQCD calculation for pp collisions: the ratio is around 0.5, and moves closer to unity when more peripheral collisions are analysed, as expected for a quenching in the QGP.  
The \iaa calculations from the presented theoretical models, which include energy loss, agree with the measurements in all three centrality classes.
In the lowest-\zt\ interval, the \ipqcd\ ratio tends to be higher than at larger \zt, for all three centrality classes, which is expected for central collisions. 
The \ipqcd\ is also compared to an NLO pQCD \iaa\ prediction that only includes cold nuclear matter effects (CNM) and no energy loss by using EPPS21 nuclear parton distribution functions (nPDFs)~\cite{Eskola:2021nhw}. 
Uncertainties related to the nPDFs are given at 90\% confidence level and were obtained by performing the calculations for each of the 107 eigenvector sets of EPPS21, resulting in uncertainties of the order of 20\%. The uncertainty relative to the factorisation scale $\mu$ is also shown, and it largely cancels out in the ratio with values around 1.5\%.
This \iaa\ prediction is provided without centrality dependence. The calculation is close to unity and cannot reproduce the observed magnitude of the suppression observed in central \PbPb\ collisions, unlike the other models that include energy loss.

\begin{figure}[!htbp]
    \begin{center}
    \includegraphics[width=0.62\textwidth]{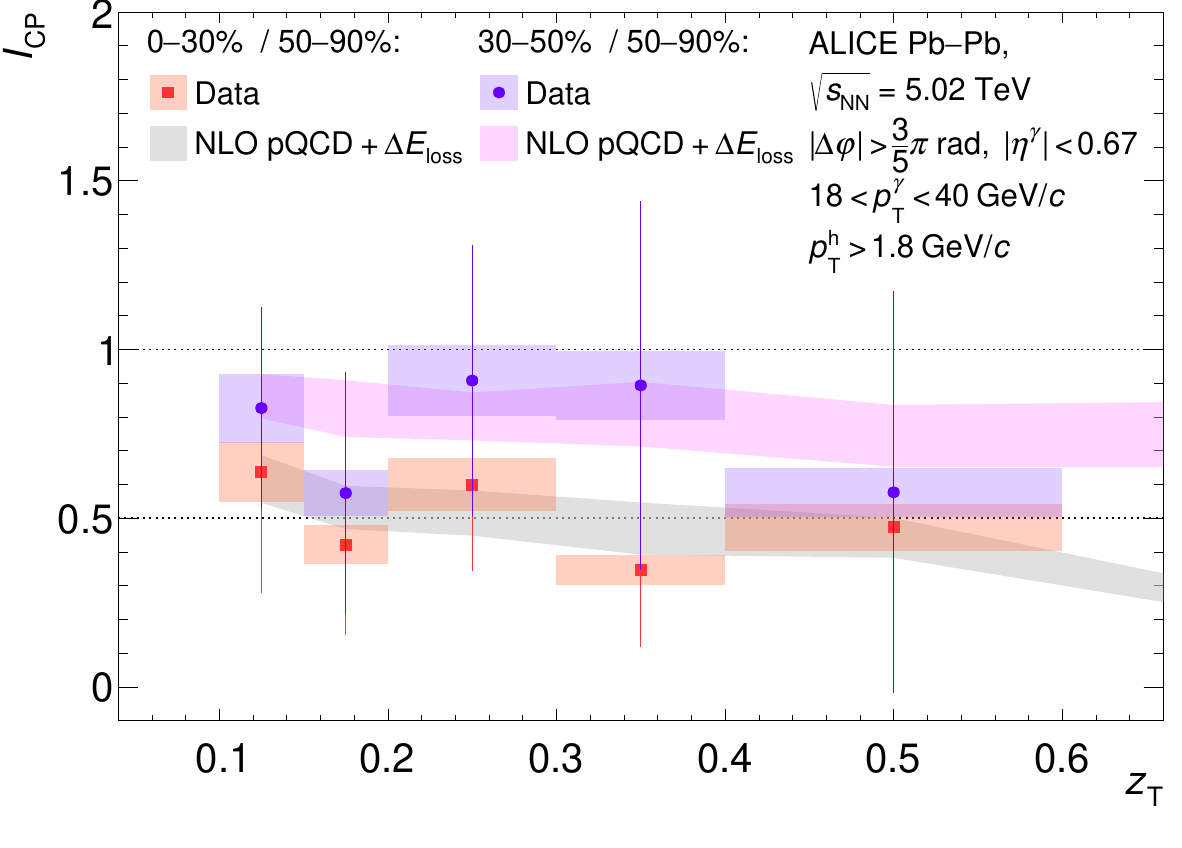}
    \end{center}
    \caption{\label{fig:Icp}(colour online) 
 $I_{\rm CP}$ ratio of the D($z_{\rm T}$) distributions for Pb$-$Pb collisions at $\sqrt{s_{\mathrm{NN}}}~=~5.02$~Te\kern-.1emV in Fig.~\ref{fig:Dzt}  for data: 0$-$30\% over  50$-$90\% (red squares), and 30$-$50\% over 50$-$90\% (violet bullets). 
 The boxes represent the systematic uncertainties, while the vertical bars indicate the statistical uncertainties. 
 Also, the equivalent ratios from the NLO pQCD calculation for Pb$-$Pb collisions, including energy loss, are shown as a band indicating the theory uncertainty.}
\end{figure}

Another interesting ratio is the \icp\ defined as the ratio between the \Dzt\ distributions in central or semicentral collisions over the peripheral ones,
  $\icp = \frac{\Dzt_{\rm{0-30\%,~30-50\%}}}{\Dzt_{\rm{50-90\%}}}$.
Figure~\ref{fig:Icp} shows the measured \icp\ as a function of \zt\ for central and semicentral centralities. 
Statistical uncertainties also dominate the results, but the expected dependence of the suppression on centrality is well visible.
The distributions are rather flat and, on average, of the order of 0.5 and 0.75, respectively.
The results are compared to the equivalent theory ratio calculated using the NLO pQCD+$\Delta E_{\rm loss}$ model, showing agreement. \\

\begin{figure}[!hbtp]
    \begin{center}
      \includegraphics[width=0.62\textwidth]{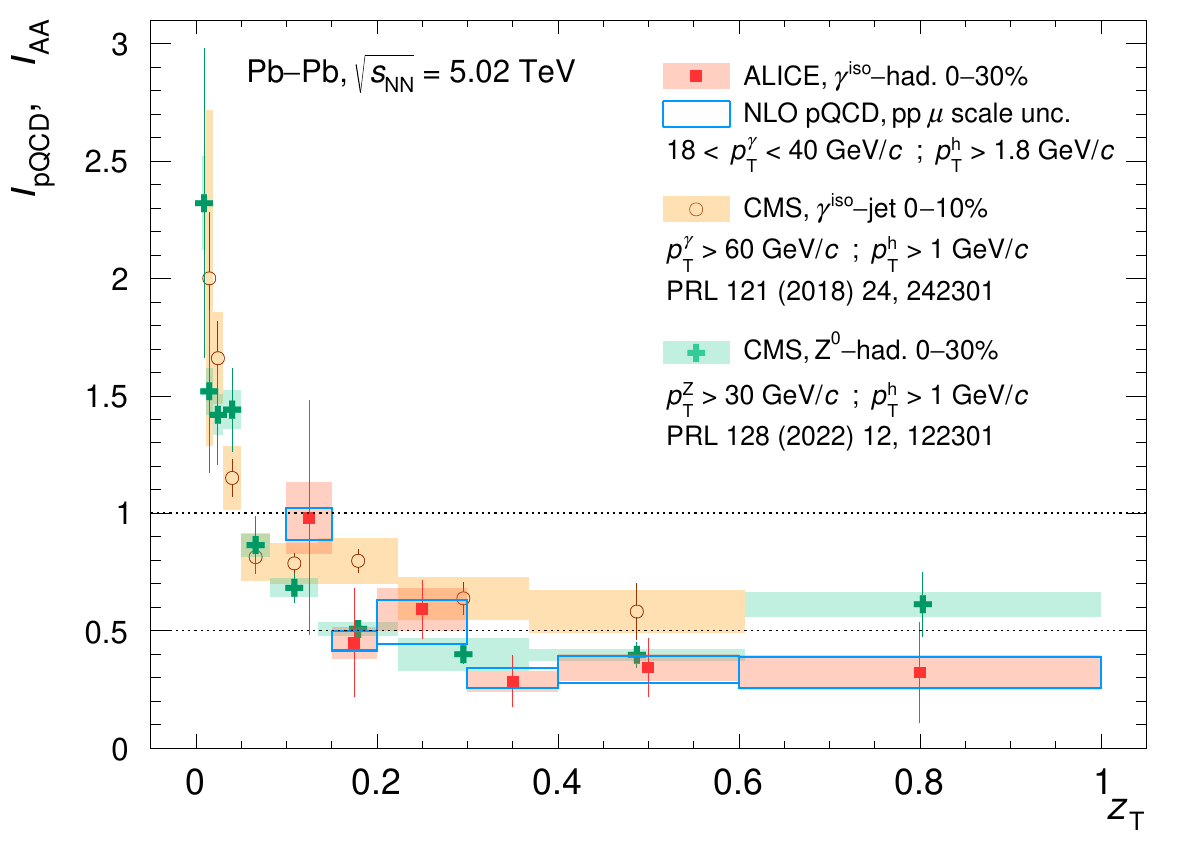}
    \end{center}
    \caption{\label{fig:CorrLHC}(colour online) 
$I_{\rm pQCD}$ for central Pb$-$Pb collisions at $\sqrt{s_{\mathrm{NN}}}~=~5.02$~Te\kern-.1emV measured in ALICE for isolated-prompt $\gamma-$hadron correlations, and $I_{\rm AA}$ measured in CMS for isolated-prompt $\gamma-$jet correlations~\cite{CMS:2018mqn} and Z$^{0}-$hadron correlations~\cite{CMS:2021otx} also in central collisions. 
 The boxes and vertical lines represent the systematic and statistical uncertainties, respectively. 
 For ALICE, the blue-open boxes represent the $\mu$ scale uncertainty for the denominator NLO pQCD calculation for pp collisions. 
}
\end{figure}

\begin{figure}[hbtp]
    \begin{center}
      \includegraphics[width=0.62\textwidth]{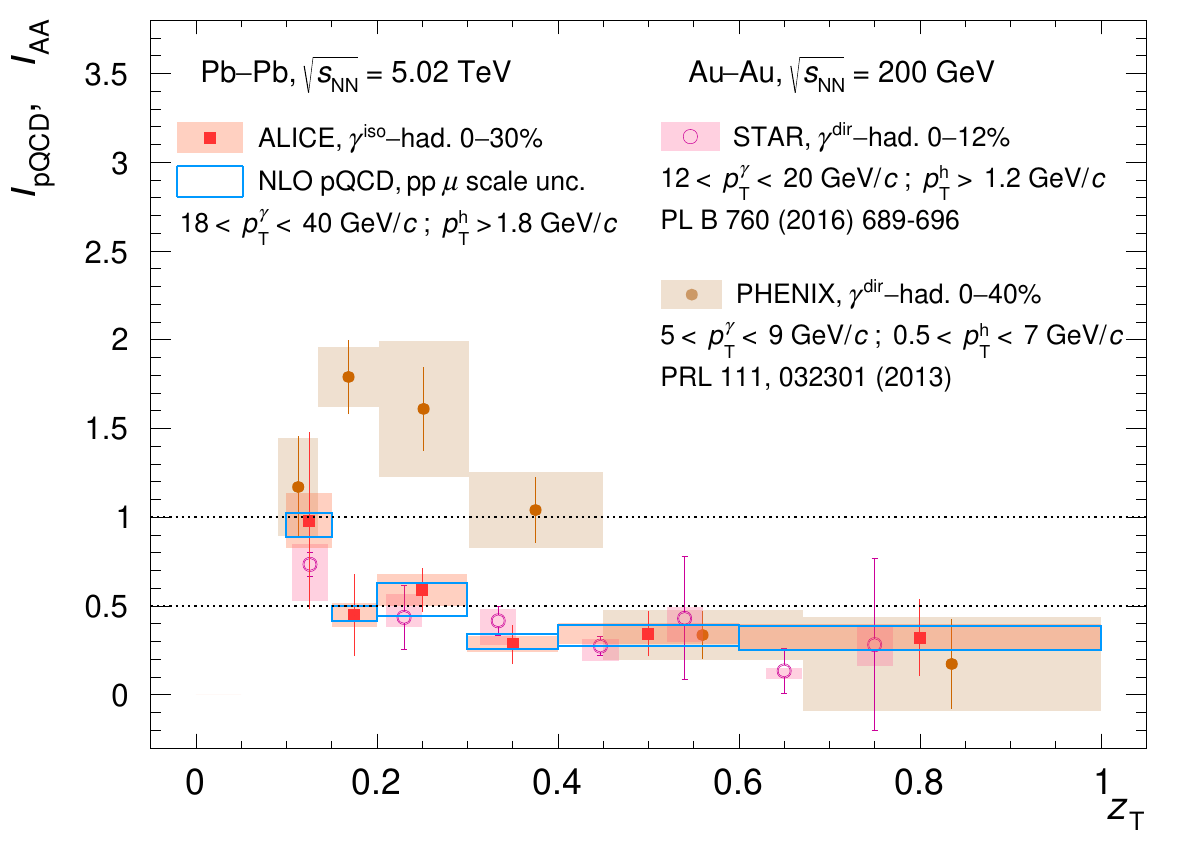}
    \end{center}
    \caption{\label{fig:CorrRHIC}(colour online) 
Isolated-prompt $\gamma-$hadron correlations \ipqcd\ for central \PbPb collisions at $\sqrt{s_{\mathrm{NN}}}~=~5.02$~Te\kern-.1emV measured by the ALICE Collaboration, and  \iaa\ for central Au$-$Au collisions at $\sqrt{s_{\rm NN}} = 200$~Ge\kern-.1emV measured by the PHENIX~\cite{PHENIX:2012aba} and STAR~\cite{STAR:2016jdz} Collaborations at RHIC, also in central collisions. 
The boxes represent the systematic uncertainties, while the vertical bars indicate the statistical uncertainties. 
For ALICE, the blue-open boxes represent the $\mu$ scale uncertainty for the denominator NLO pQCD calculation for pp collisions.
}
\end{figure}

In Figs.~\ref{fig:CorrLHC} and~\ref{fig:CorrRHIC}, these \ipqcd\ results are compared with other equivalent \iaa\ measurements: 
respectively isolated $\gamma-$jet~\cite{CMS:2018mqn} and Z$^{0}-$hadron~\cite{CMS:2021otx} correlations in \PbPb\ collisions at \snnfive\ at the LHC by the CMS Collaboration; 
and direct $\gamma-$hadron correlations obtained by the STAR~\cite{STAR:2016jdz} and PHENIX~\cite{PHENIX:2012aba} Collaborations in \AuAu collisions at $\snn=200$~Ge\kern-.1emV at RHIC.

Since the \pt\ and rapidity ranges of the trigger and associated hadrons, and the centrality ranges used by ALICE and the other experiments are not the same (see Figs.~\ref{fig:CorrLHC} and~\ref{fig:CorrRHIC} for details), and ALICE uses a theoretical pp reference, they can only be compared qualitatively.
The ALICE Collaboration has lower \ptg than the CMS Collaboration measurements (18 to 40~\GeVc compared to \ptg\ above 60~\GeVc or $\pt^{\rm \zz}$ above 30~\GeVc). 
The three distributions are compatible in the common \zt\ ranges. At low \zt\ (below 0.15), the CMS isolated $\gamma-$jet and Z$^{0}-$hadron correlations can probe the enhancement of the low-\pt\ hadrons. 
Overall, the comparison hints at a better agreement 
between the ALICE and CMS Z$^{0}-$hadron correlation since they have the same centrality range and closer trigger \pt.


The STAR experiment also reports a similar \iaa\ measurement, although in a lower \ptg\ region (12 to 20~\GeVc) and in more central collisions than ALICE. 
Their results are compatible and overlap in the common \zt~range, but the PHENIX \iaa differs significantly below \zt~=~0.45. 
The higher \zt\ yield reported by the PHENIX Collaboration can be explained by the lower \ptg: the lower energy-associated partons lose more energy relatively. Also, the measurement could be more sensitive to the enhancement of soft hadrons: 
the lowest \ptH at \zt~=~0.1 is 0.5~\GeVc for PHENIX, while it is 1.2 and 1.8~\GeVc for STAR and  ALICE, respectively. 
Also, the higher \zt\ yield reported by the PHENIX Collaboration
can reflect the higher relative energy loss of the lower energy of the associated partons.
Despite the differences in the various measurements, the trends and magnitudes 
of the \iaa\ and the \ipqcd\ ratios in the common \zt~ranges measured at LHC and RHIC are similar, except for PHENIX due to the low trigger momentum.

\section{Conclusions\label{sec:conclusion}}

The isolated-prompt photon$-$hadron correlation in \PbPb collisions at \snnfive was measured by the ALICE experiment for different centrality classes, in the transverse-momentum and pseudorapidity ranges $18<\ptg<40$~\GeVc and $|\etag| <0.67$ for the photon triggers, and $\ptH>1.8$~\GeVc and $|\etaH| <0.9$ for the associated charged particles.  
The measured \Dzt conditional yields show an agreement with NLO pQCD+$\Delta E_{\rm loss}$ and ColBT-hydro calculations that both include energy-loss mechanisms.  The \ipqcd\ ratio, approximately equivalent to the \iaa, replacing the reference measured in \pp collisions data with an NLO pQCD calculation without energy loss (in agreement with \Dzt measurements in pp collisions at lower \ptg), shows a strong suppression in central collisions and less suppression in semicentral collisions.  
The uncertainties do not allow declaring that a suppression is also visible in peripheral collisions. In all three centrality intervals, the models including jet quenching agree with the data. 
At the same time, NLO pQCD calculations, which include only cold-nuclear-matter effects, fail to reproduce the measured \ipqcd\ ratio for central \PbPb collisions. 
The \icp ratio shows a decrease in \Dzt\ when going from peripheral to more central collisions, and also agrees with NLO pQCD+$\Delta E_{\rm loss}$ calculations.

The \ipqcd\ results for central \PbPb collisions have been compared with \iaa measurements from CMS and STAR Collaborations. 
Although they use different selection criteria, \snn, and different pp collisions reference, the trends in the same \zt\ intervals agree and are consistent with a large suppression of high-\pt\ hadrons at high \zt in the most central Pb--Pb collisions.
The current measurement extends the lower limit of \ptg\ to a lower value compared to previous measurements by other LHC experiments. 
The results have also been compared with PHENIX Collaboration results, which use a lower trigger and hadron \pt than STAR and ALICE, finding a significantly enhanced soft hadron distribution at low \zt, likely due to the larger relative energy loss of the back-to-back lower-energy partons.

This measurement serves as a benchmark for Run~3 and upcoming Run~4 isolated-prompt photon$-$hadron correlation analyses in ALICE. 
With new datasets offering increasingly robust statistical samples, particularly the pp collisions sample at the same centre-of-mass energy, more accurate correlation measurements will be possible, allowing for a more precise investigation of different centrality classes and access to lower and higher \ptg\ values and smaller \zt\ intervals. 
The increased statistics will enable differential studies, such as examining correlations in different event-plane angle regions in the \PbPb collisions 30$-$50\% centrality class. 


\newenvironment{acknowledgement}{\relax}{\relax}
\begin{acknowledgement}
\section*{Acknowledgements}

The authors would like to thank Xin-Nian Wang and the CCNU theory group, particularly Man Xie and Zhong Yang, for providing the NLO pQCD (with and without energy-loss and cold-nuclear-matter effects) and ColBT-hydro calculations, respectively.


The ALICE Collaboration would like to thank all its engineers and technicians for their invaluable contributions to the construction of the experiment and the CERN accelerator teams for the outstanding performance of the LHC complex.
The ALICE Collaboration gratefully acknowledges the resources and support provided by all Grid centres and the Worldwide LHC Computing Grid (WLCG) collaboration.
The ALICE Collaboration acknowledges the following funding agencies for their support in building and running the ALICE detector:
A. I. Alikhanyan National Science Laboratory (Yerevan Physics Institute) Foundation (ANSL), State Committee of Science and World Federation of Scientists (WFS), Armenia;
Austrian Academy of Sciences, Austrian Science Fund (FWF): [M 2467-N36] and Nationalstiftung f\"{u}r Forschung, Technologie und Entwicklung, Austria;
Ministry of Communications and High Technologies, National Nuclear Research Center, Azerbaijan;
Rede Nacional de Física de Altas Energias (Renafae), Financiadora de Estudos e Projetos (Finep), Funda\c{c}\~{a}o de Amparo \`{a} Pesquisa do Estado de S\~{a}o Paulo (FAPESP) and The Sao Paulo Research Foundation  (FAPESP), Brazil;
Bulgarian Ministry of Education and Science, within the National Roadmap for Research Infrastructures 2020-2027 (object CERN), Bulgaria;
Ministry of Education of China (MOEC) , Ministry of Science \& Technology of China (MSTC) and National Natural Science Foundation of China (NSFC), China;
Ministry of Science and Education and Croatian Science Foundation, Croatia;
Centro de Aplicaciones Tecnol\'{o}gicas y Desarrollo Nuclear (CEADEN), Cubaenerg\'{\i}a, Cuba;
Ministry of Education, Youth and Sports of the Czech Republic, Czech Republic;
The Danish Council for Independent Research | Natural Sciences, the VILLUM FONDEN and Danish National Research Foundation (DNRF), Denmark;
Helsinki Institute of Physics (HIP), Finland;
Commissariat \`{a} l'Energie Atomique (CEA) and Institut National de Physique Nucl\'{e}aire et de Physique des Particules (IN2P3) and Centre National de la Recherche Scientifique (CNRS), France;
Bundesministerium f\"{u}r Forschung, Technologie und Raumfahrt (BMFTR) and GSI Helmholtzzentrum f\"{u}r Schwerionenforschung GmbH, Germany;
National Research, Development and Innovation Office, Hungary;
Department of Atomic Energy Government of India (DAE), Department of Science and Technology, Government of India (DST), University Grants Commission, Government of India (UGC) and Council of Scientific and Industrial Research (CSIR), India;
National Research and Innovation Agency - BRIN, Indonesia;
Istituto Nazionale di Fisica Nucleare (INFN), Italy;
Japanese Ministry of Education, Culture, Sports, Science and Technology (MEXT) and Japan Society for the Promotion of Science (JSPS) KAKENHI, Japan;
Consejo Nacional de Ciencia (CONACYT) y Tecnolog\'{i}a, through Fondo de Cooperaci\'{o}n Internacional en Ciencia y Tecnolog\'{i}a (FONCICYT) and Direcci\'{o}n General de Asuntos del Personal Academico (DGAPA), Mexico;
Nederlandse Organisatie voor Wetenschappelijk Onderzoek (NWO), Netherlands;
The Research Council of Norway, Norway;
Pontificia Universidad Cat\'{o}lica del Per\'{u}, Peru;
Ministry of Science and Higher Education, National Science Centre and WUT ID-UB, Poland;
Korea Institute of Science and Technology Information and National Research Foundation of Korea (NRF), Republic of Korea;
Ministry of Education and Scientific Research, Institute of Atomic Physics, Ministry of Research and Innovation and Institute of Atomic Physics and Universitatea Nationala de Stiinta si Tehnologie Politehnica Bucuresti, Romania;
Ministerstvo skolstva, vyskumu, vyvoja a mladeze SR, Slovakia;
National Research Foundation of South Africa, South Africa;
Swedish Research Council (VR) and Knut \& Alice Wallenberg Foundation (KAW), Sweden;
European Organization for Nuclear Research, Switzerland;
Suranaree University of Technology (SUT), National Science and Technology Development Agency (NSTDA) and National Science, Research and Innovation Fund (NSRF via PMU-B B05F650021), Thailand;
Turkish Energy, Nuclear and Mineral Research Agency (TENMAK), Turkey;
National Academy of  Sciences of Ukraine, Ukraine;
Science and Technology Facilities Council (STFC), United Kingdom;
National Science Foundation of the United States of America (NSF) and United States Department of Energy, Office of Nuclear Physics (DOE NP), United States of America.
In addition, individual groups or members have received support from:
Czech Science Foundation (grant no. 23-07499S), Czech Republic;
FORTE project, reg.\ no.\ CZ.02.01.01/00/22\_008/0004632, Czech Republic, co-funded by the European Union, Czech Republic;
European Research Council (grant nos. 101220549, 950692), European Union;
Deutsche Forschungs Gemeinschaft (DFG, German Research Foundation) ``Neutrinos and Dark Matter in Astro- and Particle Physics'' (grant no. SFB 1258), Germany;
CONVECS project, CUP C97H23001700002 FESR 2021-2027 program, Italy.

\end{acknowledgement}

\bibliographystyle{utphys}   
\bibliography{bibliography}

\newpage

\appendix

\section{The ALICE Collaboration}
\label{app:collab}
\begin{flushleft} 
\small

D.A.H.~Abdallah\,\orcidlink{0000-0003-4768-2718}\,$^{\rm 134}$, 
I.J.~Abualrob\,\orcidlink{0009-0005-3519-5631}\,$^{\rm 112}$, 
S.~Acharya\,\orcidlink{0000-0002-9213-5329}\,$^{\rm 49}$, 
K.~Agarwal\,\orcidlink{0000-0001-5781-3393}\,$^{\rm II,}$$^{\rm 23}$, 
G.~Aglieri Rinella\,\orcidlink{0000-0002-9611-3696}\,$^{\rm 32}$, 
L.~Aglietta\,\orcidlink{0009-0003-0763-6802}\,$^{\rm 24}$, 
N.~Agrawal\,\orcidlink{0000-0003-0348-9836}\,$^{\rm 25}$, 
Z.~Ahammed\,\orcidlink{0000-0001-5241-7412}\,$^{\rm 132}$, 
S.~Ahmad\,\orcidlink{0000-0003-0497-5705}\,$^{\rm 15}$, 
I.~Ahuja\,\orcidlink{0000-0002-4417-1392}\,$^{\rm 36}$
% ...

\section*{Affiliation Notes}

$^{\rm I}$ Deceased\\
$^{\rm II}$ Also at: INFN Trieste, Trieste, Italy\\
$^{\rm III}$ Also at: Fondazione Bruno Kessler (FBK), Trento, Italy\\
$^{\rm IV}$ Also at: Czech Technical University in Prague, Prague, Czech Republic\\
$^{\rm V}$ Also at: Instituto de Fisica da Universidade de Sao Paulo\\
$^{\rm VI}$ Also at: Dipartimento DET del Politecnico di Torino, Turin, Italy\\
$^{\rm VII}$ Also at: University College of Dublin, Dublin, Ireland\\
$^{\rm VIII}$ Also at: Department of Applied Physics, Aligarh Muslim University, Aligarh, India\\
$^{\rm IX}$ Also at: Institute of Theoretical Physics, University of Wroclaw, Wroclaw, Poland\\
$^{\rm X}$ Also at: Facultad de Ciencias, Universidad Nacional Aut\'{o}noma de M\'{e}xico, Mexico City, Mexico\\

\section*{Collaboration Institutes}

$^{1}$ A.I. Alikhanyan National Science Laboratory (Yerevan Physics Institute) Foundation, Yerevan, Armenia\\
$^{2}$ AGH University of Krakow, Cracow, Poland\\
$^{3}$ Bogolyubov Institute for Theoretical Physics, National Academy of Sciences of Ukraine, Kyiv, Ukraine\\
$^{4}$ Bose Institute, Department of Physics  and Centre for Astroparticle Physics and Space Science (CAPSS), Kolkata, India\\
$^{5}$ California Polytechnic State University, San Luis Obispo, California, United States\\
$^{6}$ Central China Normal University, Wuhan, China\\
$^{7}$ Centro de Aplicaciones Tecnol\'{o}gicas y Desarrollo Nuclear (CEADEN), Havana, Cuba\\
$^{8}$ Centro de Investigaci\'{o}n y de Estudios Avanzados (CINVESTAV), Mexico City and M\'{e}rida, Mexico\\
$^{9}$ Chicago State University, Chicago, Illinois, United States\\
$^{10}$ China Nuclear Data Center, China Institute of Atomic Energy, Beijing, China\\
$^{11}$ China University of Geosciences, Wuhan, China\\
$^{12}$ Chungbuk National University, Cheongju, Republic of Korea\\
$^{13}$ Comenius University Bratislava, Faculty of Mathematics, Physics and Informatics, Bratislava, Slovak Republic\\
$^{14}$ Creighton University, Omaha, Nebraska, United States\\
$^{15}$ Department of Physics, Aligarh Muslim University, Aligarh, India\\
$^{16}$ Department of Physics, Pusan National University, Pusan, Republic of Korea\\
$^{17}$ Department of Physics, Sejong University, Seoul, Republic of Korea\\
$^{18}$ Department of Physics, University of California, Berkeley, California, United States\\
$^{19}$ Department of Physics, University of Oslo, Oslo, Norway\\
$^{20}$ Department of Physics and Technology, University of Bergen, Bergen, Norway\\
$^{21}$ Dipartimento di Fisica, Universit\`{a} di Pavia, Pavia, Italy\\
$^{22}$ Dipartimento di Fisica dell'Universit\`{a} and Sezione INFN, Cagliari, Italy\\
$^{23}$ Dipartimento di Fisica dell'Universit\`{a} and Sezione INFN, Trieste, Italy\\
$^{24}$ Dipartimento di Fisica dell'Universit\`{a} and Sezione INFN, Turin, Italy\\
$^{25}$ Dipartimento di Fisica e Astronomia dell'Universit\`{a} and Sezione INFN, Bologna, Italy\\
$^{26}$ Dipartimento di Fisica e Astronomia dell'Universit\`{a} and Sezione INFN, Catania, Italy\\
$^{27}$ Dipartimento di Fisica e Astronomia dell'Universit\`{a} and Sezione INFN, Padova, Italy\\
$^{28}$ Dipartimento di Fisica `E.R.~Caianiello' dell'Universit\`{a} and Gruppo Collegato INFN, Salerno, Italy\\
$^{29}$ Dipartimento DISAT del Politecnico and Sezione INFN, Turin, Italy\\
$^{30}$ Dipartimento di Scienze MIFT, Universit\`{a} di Messina, Messina, Italy\\
$^{31}$ Dipartimento Interateneo di Fisica `M.~Merlin' and Sezione INFN, Bari, Italy\\
$^{32}$ European Organization for Nuclear Research (CERN), Geneva, Switzerland\\
$^{33}$ Faculty of Electrical Engineering, Mechanical Engineering and Naval Architecture, University of Split, Split, Croatia\\
$^{34}$ Faculty of Nuclear Sciences and Physical Engineering, Czech Technical University in Prague, Prague, Czech Republic\\
$^{35}$ Faculty of Physics, Sofia University, Sofia, Bulgaria\\
$^{36}$ Faculty of Science, P.J.~\v{S}af\'{a}rik University, Ko\v{s}ice, Slovak Republic\\
$^{37}$ Faculty of Technology, Environmental and Social Sciences, Bergen, Norway\\
$^{38}$ Frankfurt Institute for Advanced Studies, Johann Wolfgang Goethe-Universit\"{a}t Frankfurt, Frankfurt, Germany\\
$^{39}$ Fudan University, Shanghai, China\\
$^{40}$ Gauhati University, Department of Physics, Guwahati, India\\
$^{41}$ Helmholtz-Institut f\"{u}r Strahlen- und Kernphysik, Rheinische Friedrich-Wilhelms-Universit\"{a}t Bonn, Bonn, Germany\\
$^{42}$ Helsinki Institute of Physics (HIP), Helsinki, Finland\\
$^{43}$ High Energy Physics Group,  Universidad Aut\'{o}noma de Puebla, Puebla, Mexico\\
$^{44}$ Horia Hulubei National Institute of Physics and Nuclear Engineering, Bucharest, Romania\\
$^{45}$ HUN-REN Wigner Research Centre for Physics, Budapest, Hungary\\
$^{46}$ Indian Institute of Technology Bombay (IIT), Mumbai, India\\
$^{47}$ Indian Institute of Technology Indore, Indore, India\\
$^{48}$ INFN, Laboratori Nazionali di Frascati, Frascati, Italy\\
$^{49}$ INFN, Sezione di Bari, Bari, Italy\\
$^{50}$ INFN, Sezione di Bologna, Bologna, Italy\\
$^{51}$ INFN, Sezione di Cagliari, Cagliari, Italy\\
$^{52}$ INFN, Sezione di Catania, Catania, Italy\\
$^{53}$ INFN, Sezione di Padova, Padova, Italy\\
$^{54}$ INFN, Sezione di Pavia, Pavia, Italy\\
$^{55}$ INFN, Sezione di Torino, Turin, Italy\\
$^{56}$ INFN, Sezione di Trieste, Trieste, Italy\\
$^{57}$ Inha University, Incheon, Republic of Korea\\
$^{58}$ Institute for Gravitational and Subatomic Physics (GRASP), Utrecht University/Nikhef, Utrecht, Netherlands\\
$^{59}$ Institute of Experimental Physics, Slovak Academy of Sciences, Ko\v{s}ice, Slovak Republic\\
$^{60}$ Institute of Physics, Homi Bhabha National Institute, Bhubaneswar, India\\
$^{61}$ Institute of Physics of the Czech Academy of Sciences, Prague, Czech Republic\\
$^{62}$ Institute of Space Science (ISS), Bucharest, Romania\\
$^{63}$ Institut f\"{u}r Kernphysik, Johann Wolfgang Goethe-Universit\"{a}t Frankfurt, Frankfurt, Germany\\
$^{64}$ Instituto de Ciencias Nucleares, Universidad Nacional Aut\'{o}noma de M\'{e}xico, Mexico City, Mexico\\
$^{65}$ Instituto de F\'{i}sica, Universidade Federal do Rio Grande do Sul (UFRGS), Porto Alegre, Brazil\\
$^{66}$ Instituto de F\'{\i}sica, Universidad Nacional Aut\'{o}noma de M\'{e}xico, Mexico City, Mexico\\
$^{67}$ iThemba LABS, National Research Foundation, Somerset West, South Africa\\
$^{68}$ Jeonbuk National University, Jeonju, Republic of Korea\\
$^{69}$ Korea Institute of Science and Technology Information, Daejeon, Republic of Korea\\
$^{70}$ Laboratoire de Physique Subatomique et de Cosmologie, Universit\'{e} Grenoble-Alpes, CNRS-IN2P3, Grenoble, France\\
$^{71}$ Lawrence Berkeley National Laboratory, Berkeley, California, United States\\
$^{72}$ Lund University Department of Physics, Division of Particle Physics, Lund, Sweden\\
$^{73}$ Marietta Blau Institute, Vienna, Austria\\
$^{74}$ Nagasaki Institute of Applied Science, Nagasaki, Japan\\
$^{75}$ Nara Women{'}s University (NWU), Nara, Japan\\
$^{76}$ National Centre for Nuclear Research, Warsaw, Poland\\
$^{77}$ National Institute of Science Education and Research, Homi Bhabha National Institute, Jatni, India\\
$^{78}$ National Nuclear Research Center, Baku, Azerbaijan\\
$^{79}$ National Research and Innovation Agency - BRIN, Jakarta, Indonesia\\
$^{80}$ Niels Bohr Institute, University of Copenhagen, Copenhagen, Denmark\\
$^{81}$ Nikhef, National institute for subatomic physics, Amsterdam, Netherlands\\
$^{82}$ Nuclear Physics Group, STFC Daresbury Laboratory, Daresbury, United Kingdom\\
$^{83}$ Nuclear Physics Institute of the Czech Academy of Sciences, Husinec-\v{R}e\v{z}, Czech Republic\\
$^{84}$ Oak Ridge National Laboratory, Oak Ridge, Tennessee, United States\\
$^{85}$ Ohio State University, Columbus, Ohio, United States\\
$^{86}$ Physics department, Faculty of science, University of Zagreb, Zagreb, Croatia\\
$^{87}$ Physics Department, Panjab University, Chandigarh, India\\
$^{88}$ Physics Department, University of Jammu, Jammu, India\\
$^{89}$ Physics Program and International Institute for Sustainability with Knotted Chiral Meta Matter (WPI-SKCM$^{2}$), Hiroshima University, Hiroshima, Japan\\
$^{90}$ Physikalisches Institut, Eberhard-Karls-Universit\"{a}t T\"{u}bingen, T\"{u}bingen, Germany\\
$^{91}$ Physikalisches Institut, Ruprecht-Karls-Universit\"{a}t Heidelberg, Heidelberg, Germany\\
$^{92}$ Physik Department, Technische Universit\"{a}t M\"{u}nchen, Munich, Germany\\
$^{93}$ Politecnico di Bari and Sezione INFN, Bari, Italy\\
$^{94}$ Research Division and ExtreMe Matter Institute EMMI, GSI Helmholtzzentrum f\"ur Schwerionenforschung GmbH, Darmstadt, Germany\\
$^{95}$ Saga University, Saga, Japan\\
$^{96}$ Saha Institute of Nuclear Physics, Homi Bhabha National Institute, Kolkata, India\\
$^{97}$ School of Physics and Astronomy, University of Birmingham, Birmingham, United Kingdom\\
$^{98}$ Secci\'{o}n F\'{\i}sica, Departamento de Ciencias, Pontificia Universidad Cat\'{o}lica del Per\'{u}, Lima, Peru\\
$^{99}$ SUBATECH, IMT Atlantique, Nantes Universit\'{e}, CNRS-IN2P3, Nantes, France\\
$^{100}$ Sungkyunkwan University, Suwon City, Republic of Korea\\
$^{101}$ Suranaree University of Technology, Nakhon Ratchasima, Thailand\\
$^{102}$ Technical University of Ko\v{s}ice, Ko\v{s}ice, Slovak Republic\\
$^{103}$ The Henryk Niewodniczanski Institute of Nuclear Physics, Polish Academy of Sciences, Cracow, Poland\\
$^{104}$ The University of Texas at Austin, Austin, Texas, United States\\
$^{105}$ Universidad Aut\'{o}noma de Sinaloa, Culiac\'{a}n, Mexico\\
$^{106}$ Universidade de S\~{a}o Paulo (USP), S\~{a}o Paulo, Brazil\\
$^{107}$ Universidade Estadual de Campinas (UNICAMP), Campinas, Brazil\\
$^{108}$ Universidade Federal do ABC, Santo Andre, Brazil\\
$^{109}$ Universitatea Nationala de Stiinta si Tehnologie Politehnica Bucuresti, Bucharest, Romania\\
$^{110}$ University of Cape Town, Cape Town, South Africa\\
$^{111}$ University of Derby, Derby, United Kingdom\\
$^{112}$ University of Houston, Houston, Texas, United States\\
$^{113}$ University of Jyv\"{a}skyl\"{a}, Jyv\"{a}skyl\"{a}, Finland\\
$^{114}$ University of Kansas, Lawrence, Kansas, United States\\
$^{115}$ University of Liverpool, Liverpool, United Kingdom\\
$^{116}$ University of Science and Technology of China, Hefei, China\\
$^{117}$ University of Silesia in Katowice, Katowice, Poland\\
$^{118}$ University of South-Eastern Norway, Kongsberg, Norway\\
$^{119}$ University of Tennessee, Knoxville, Tennessee, United States\\
$^{120}$ University of the Witwatersrand, Johannesburg, South Africa\\
$^{121}$ University of Tokyo, Tokyo, Japan\\
$^{122}$ University of Tsukuba, Tsukuba, Japan\\
$^{123}$ Universit\"{a}t M\"{u}nster, Institut f\"{u}r Kernphysik, M\"{u}nster, Germany\\
$^{124}$ Universit\'{e} Clermont Auvergne, CNRS/IN2P3, LPC, Clermont-Ferrand, France\\
$^{125}$ Universit\'{e} de Lyon, CNRS/IN2P3, Institut de Physique des 2 Infinis de Lyon, Lyon, France\\
$^{126}$ Universit\'{e} de Strasbourg, CNRS, IPHC UMR 7178, F-67000 Strasbourg, France, Strasbourg, France\\
$^{127}$ Universit\'{e} Paris-Saclay, Centre d'Etudes de Saclay (CEA), IRFU, D\'{e}partment de Physique Nucl\'{e}aire (DPhN), Saclay, France\\
$^{128}$ Universit\'{e}  Paris-Saclay, CNRS/IN2P3, IJCLab, Orsay, France\\
$^{129}$ Universit\`{a} degli Studi di Foggia, Foggia, Italy\\
$^{130}$ Universit\`{a} del Piemonte Orientale, Vercelli, Italy\\
$^{131}$ Universit\`{a} di Brescia, Brescia, Italy\\
$^{132}$ Variable Energy Cyclotron Centre, Homi Bhabha National Institute, Kolkata, India\\
$^{133}$ Warsaw University of Technology, Warsaw, Poland\\
$^{134}$ Wayne State University, Detroit, Michigan, United States\\
$^{135}$ Yale University, New Haven, Connecticut, United States\\
$^{136}$ Yildiz Technical University, Istanbul, Turkey\\
$^{137}$ Yonsei University, Seoul, Republic of Korea\\
$^{138}$ Affiliated with an institute formerly covered by a cooperation agreement with CERN\\
$^{139}$ Affiliated with an international laboratory covered by a cooperation agreement with CERN.\\

\end{flushleft} 
  
\end{document}